\documentclass[10pt]{article}
\date{}
\title{Qualms regarding \lq\lq Dynamical Foundations of Nonextensive Statistical
Mechanics \rq\rq\
by C. Beck (cond-mat/0105374)}
   \author{B. H. Lavenda$^1$ and J. Dunning-Davies$^2$\\
$^1$Universit\`a degli Studi  Camerino 62032 (MC) Italy;\\ email: bernard.lavenda@unicam.it\\
$^2$ Department of Physics, University of Hull, Hull HU6
7RX\\ England; email: j.dunning-davies@hull.ac.uk}
\usepackage{eufrak}
\newcommand{\half}{\mbox{\small$\frac{1}{2}$}}

\newcommand{\sumn}{\sum_{i=1}^n\,}

\begin{document}
\maketitle
\begin{abstract} 
The derivation of Student's pdf from superstatistics is a mere coincidence due to the
choice of the $\chi^2$ distribution for the inverse temperature which is actually the
Maxwell distribution for the speed. The difference between the estimator and the variance 
introduces a fluctuating temperature that is generally different than the temperature
of the variance. Fluctuations in the energy of a composite system are transferred to
fluctuations in the temperature via L\'evy's transform. The randomization of the operational parameter 
leading to a subordinated process requires the same number of degrees of freedom in the
directing as in the original process. In place of Student's pdf we find  the
subordinated process to Maxwell's speed distribution has a Pareto distribution
which is interpreted as the density of the length of the random
velocity vector.
\end{abstract}
\flushbottom
There is always a certain delight to see old ideas from a new perspective, but there is 
no delight to see professed \lq new\rq\ ideas that are not new. The notion of \lq superstatistics\rq 
\cite{BC} grew out of an idea that if one began with a stationary Maxwell distribution
and introduced a \emph{randomized operational temperature\/}, then a new process could be
derived which has something to do with Tsallis' distribution \cite{Beck}, which he obtained as a stationary 
condition from maximizing a nonadditive entropy of degree-$q$ with respect to the
energy constraint \cite{Souza}. The generalization to \lq superstatistics\rq\ then consisted in 
replacing the exponential in the Maxwell distribution by $e^{-\beta H}$, where $H$ is any
generic hamiltonian, and using an arbitrary probability distribution for the conjugate
$\beta$ one could obtain a new probability distribution for $H$ by integrating the product of
the two probability distributions over all values of $\beta$. We have questioned the
physical justification of this procedure elsewhere \cite{LDD}.\par
The idea of randomizing temperature is not new, as Beck would have us believe; 
it can be found in ref. \cite{Lavenda95}. It is based on  subordination, which can 
be found in Feller's book \cite{Feller}. If $X(t)$ represents a brownian motion then by
\emph{randomizing the operational time\/} $t$, a variety of new processes can be 
derived. That is, if the displacements of brownian motion are governed by the
probability density function (pdf),
\begin{equation}
f_{t_0}(x)=\frac{1}{\sqrt{2\pi t_0}}e^{-x^2/2t_0}, \label{eq:brown}
\end{equation}
where $x$ is the random variable and $t_0$ the parameter, then exchanging their roles
leads to a L\'evy pdf, 
\begin{equation}
f_{x_0}(t)=
\frac{x_0}{\sqrt{2\pi t^3}}e^{-x_0^2/2t}, \label{eq:levy}
\end{equation}
for the randomized time, $t$, where $x_0$ is the parameter.
 The two pdf are related by the transform
$x^2/t_0=x_0^2/t$. The Cauchy process is then said to be subordinated to 
brownian motion \cite{Feller}
\[
f_{x_0}(x)=\int_{-\infty}^{+\infty}\,f_{t}(x)f_{x_0}(t)\,dt=\frac{x_0}{2\pi}\int_{-\infty}^{+\infty}\,\frac{1}{t^2}
e^{-(x^2+x_0^2)/2t}\,dt=\frac{1}{\pi}\frac{x_0}{x_0^2+x^2}, \]
where the two processes are evaluated at the same time.
\par The Maxwell speed, $U(T)$, in a single dimension, and the temperature, $T$, stand in 
the same relation as the displacement of a brownian motion process, $X(t)$, and
the time, $t$ \cite{Lavenda95}. The Cauchy distribution for the speed is therefore
subordinated to the Maxwell distribution in one dimension \cite{Lavenda95}.\par
Therefore, Beck \cite{Beck} does not have to suppose that the fluctuations in inverse
temperature, $\beta$ follow a $\chi^2$ distribution given by his equation (2), 
\begin{equation}
f_{\beta_0}(\beta)=\frac{1}{\Gamma(n/2)}\left(\frac{n}{2\beta_0}\right)^{n/2}\beta^{n/2-1}
e^{-n\beta/2\beta_0}, \label{eq:beck}
\end{equation}
which we would prefer to write as
\begin{equation}
f_{u_0}(\beta)=\frac{1}{\Gamma(n/2)}\left(\frac{u_0^2}{2}\right)^{n/2}
\beta^{n/2-1} e^{-\half\beta u_0^2}, \label{eq:chi}
\end{equation}
on the strength of equipartition, $u_0^2\beta_0=n$.
By what we shall show is an incorrect association of Tsallis' distribution with the
subordinated process, Beck finds $n=2/(q-1)$, where $q$ is the degree of the nonadditive
entropy, or, what is commonly known as the Tsallis exponent. A single mode of
(\ref{eq:beck}) exists at $\beta=(1-2/n)\beta_0$. The single mode therefore vanishes for
all $q\ge2$. We should expect a qualitative change in the behavior of the process for 
$q<2$ and $q\ge2$. That no behavioral change has been predicted, already casts doubts on 
the association of half the number of the degrees of freedom with the inverse of 
$(q-1)$.
\par
It follows from the L\'evy transform
\begin{equation}
\beta u_0^2=\beta_0u^2 \label{eq:levy-t}
\end{equation}
that the $n$-dimensional Maxwell speed distribution 
\begin{equation}
f_{\beta_0}(u)=\frac{2}{\Gamma(n/2)}\left(\frac{\beta_0}{2}\right)^{n/2}
u^{n-1}e^{-\half\beta_0 u^2}, \label{eq:n-maxwell}
\end{equation}
corresponds to the $\chi^2$ distribution for inverse temperature, (\ref{eq:chi}), 
and not Beck's one dimensional Maxwell pdf (6)[eqn. (\ref{eq:1-maxwell}) below]. What has happened is that, in a space of
$n$ dimensions, we began with a Markov process, $U(\beta)$, whose stationary transition
probability is given by (\ref{eq:n-maxwell}). A whole host of new processes can be
derived by randomizing the temperature, or its inverse, $\beta>0$, 
corresponding to a new random
variable $\mathfrak{B}(\beta)$ whose pdf is (\ref{eq:chi}). The process 
$\mathfrak{B}(\beta)$ is called the
\lq directing\rq\ process \cite[p. 347]{Feller}\cite[p. 56]{Lavenda95}. The pdf of
 the new process $U(\mathfrak{B}(\beta))$
will be given by
\begin{equation}
f_{u_0}(u)=\int_0^\infty\,f_{\beta}(u)f_{u_0}(\beta)\,d\beta=
\frac{2}{B(n/2,n/2)}\frac{(u/u_0)^{n-1}}{(1+(u/u_0)^2)^n}\frac{1}{u_0},
\label{eq:in-beta}
\end{equation}
where we have set $\beta=\beta_0$ so as to ensure that the two processes are in thermal 
equilibium, and $B(\cdot,\cdot)$ is the beta function. 
Contrary to what has been claimed \cite{Beck}, (\ref{eq:in-beta}) is
\emph{not\/} Tsallis' canonical probability distribution \cite[eqn (9)]{Beck}
\begin{equation}
f_{\beta_0}(u)=\frac{\Gamma\left(\frac{n+1}{2}\right)}{\sqrt{n/2}\;\Gamma(n/2)}
\cdot\sqrt{\frac{\beta_0}{2\pi}}\left(1+\frac{\beta_0u^2}{n}\right)^{-(n+1)/2}. 
\label{eq:Tsallis}
\end{equation}
From symmetry considerations, all odd order moments of (\ref{eq:Tsallis}) are zero. In particular,
the second moment is
\begin{equation}
\overline{u^2}=\frac{n\beta_0^{-1}}{n-2}, \label{eq:equi}
\end{equation}
which does not reflect  equipartition   that follows from 
(\ref{eq:n-maxwell}) at all [\emph{cf\/}., eqn (\ref{eq:equi-bis}) below ].\par
For a generic value of $n$, 
 (\ref{eq:in-beta}) is a special case of the
 inverse beta pdf
\begin{equation}
f(x)=\frac{1}{B(n/2,n/2)}\frac{x^{n/2-1}}{(1+x)^n},\label{eq:pareto}
\end{equation}
where $x=(u/u_0)^{\half}$.  In economics, (\ref{eq:pareto}) is known as the Pareto pdf, 
since \lq\lq it was thought
(rather naively from a modern statistical standpoint) that income distributions
should have a tail with a density $\sim Ax^{-\alpha}$ as $x\rightarrow\infty$, and
(\ref{eq:pareto}) fulfills this requirement\rq\rq \cite[p. 50]{Feller}.\par
In $n=1$ dimension, (\ref{eq:in-beta}) is the half-Cauchy pdf for a positive
variate. \emph{Just as the Cauchy process for the displacement is subordinated 
to brownian motion in one dimension when time is randomized,
the Cauchy process for the kinetic energy is  subordinated to the Maxwellian 
in a single dimension when the temperature, or its inverse, is randomized\/} 
\cite{Lavenda95}.\par 
In $n=3$ dimensions, (\ref{eq:in-beta}) is \cite[p. 95]{Lavenda95}
\begin{equation}
f_{n=3}(v)=\frac{8}{\pi}\frac{v^2}{(1+v^2)^3}. \label{eq:lavenda}
\end{equation}This is the density of the length of a random velocity vector 
$v=\sqrt{v_x^2+v_y^2+v_z^2}$ in $\Re^3$ dimensions. It is
related to the density 
\begin{equation}
f_{n=1}(v)=\frac{1}{B(\half;m-\half)}\frac{1}{(1+v^2)^m}, \label{eq:student}
\end{equation}
 in a fixed direction, by \cite[p. 32]{Feller}
\begin{equation}
f_{n=3}(v)=-vf^{\prime}_{n=1}(v), \label{eq:feller}
\end{equation}
where the prime denotes differentiation with respect to the argument.
The pdf (\ref{eq:student}) has $2m-1$ degrees
of freedom. For $m=1$, (\ref{eq:student}) reduces to the Cauchy pdf, while for $m=2$, 
implying that the distribution has $3$ degrees of freedom, we obtain (\ref{eq:lavenda})
from (\ref{eq:feller}). Hence, (\ref{eq:lavenda}) is the density of the length $v$ of
a random velocity vector in $n=3$ dimensions \cite[p. 95]{Lavenda95}.\par
Although (\ref{eq:beck}) is valid for any $n>0$, Beck's 
\lq conditional\rq\ pdf for the speed, (6), is our (\ref{eq:n-maxwell}) for $n=1$. For any
$n\neq1$ there is an incompatibility in the dimensions of the two distributions. This
is responsible for the seemingly close appearance of the pdf of the subordinated
process with the Tsallis distribution.  In other words, if Beck's (6) is 
meant to be the speed distribution, it does not reflect the dimensionality of the $\chi^2$
pdf for $\beta$, (\ref{eq:beck}). Rather, Beck finds that the subordinated process is 
governed by Student's $t$ distribution, (\ref{eq:Tsallis}).
The first factor tends to unity as $n\rightarrow\infty$, and for every fixed $u$
\cite{Cramer}
\[
-\frac{n+1}{2}\log\left(1+\frac{\beta_0u^2}{n}\right)\rightarrow -\half\beta_0u^2
\]
so that in the limit as $n\rightarrow\infty$ we have
\begin{equation}
f_{\beta_0}(u)=\sqrt{\frac{\beta_0}{2\pi}}e^{-\half\beta_0u^2}. \label{eq:1-maxwell}
\end{equation}
This  is the $n=1$ Maxwellian of (\ref{eq:n-maxwell}) which is the
conditional pdf (6) that Beck started out with!\par
What has happened is that the derivation of the Student pdf (\ref{eq:Tsallis}) 
\emph{formally\/} resembles subordination, but has really nothing to do with it. 
Suppose that $n+1$ random variables $u$ and $u_1,u_2,\ldots,u_n$ are independent
and identically distributed according to the normal distribution with zero mean
and standard deviation $1/\sqrt{\beta_0}$. The 
distribution of $u$ is given by (\ref{eq:1-maxwell}), while the distribution of the 
square root of the average of the sum of squares, $v=\sqrt{\frac{1}{n}
\sumn u_i^2}$ is
\begin{equation}
f_{\beta_0}(v)=\frac{2}{\Gamma(n/2)}\left(\frac{n\beta_0}{2}\right)^{n/2}
v^{n-1}e^{-\half n\beta_0 v}.
\label{eq:v}
\end{equation}
On account of the independence of the random variables $u$ and $v$, their joint pdf
will be given by
\begin{equation}
f_{\beta_0}(u)f_{\beta_0}(v)=\frac{2}{\Gamma(n/2)}
\sqrt{\frac{2}{\pi}}
\left(\frac{n}{2}\right)^{n/2}\beta_0^{(n+1)/2}v^{n-1}e^{-\half\beta_0
(u^2+nv^2)}.
\label{eq:joint}
\end{equation}
The probability that $u/v\le t$ is the integral of (\ref{eq:joint}) over the region
$v>0$ and $u\le tv$. Introducing the transformation $u=xy$ and $v=y$, whose Jacobian is
$y$, gives the distribution
\[F(t)=\frac{2}{\Gamma(n/2)}\sqrt{\frac{2}{\pi}}\left(\frac{n}{2}\right)^{n/2}
\beta_0^{(n+1)/2}
\int_{-\infty}^t\,dx\int_0^{\infty}\,dy y^ne^{-\half\-\beta_0n(1+x^2/n)y^2}.
\]
Its derivative yields Student's pdf
\begin{equation}
f(t)  = \frac{1}{\sqrt{n\pi}}\frac{\Gamma\left(\frac{n+1}{2}\right)}{\Gamma(n/2)}
\left(1+\frac{t^2}{n}\right)^{-(n+1)/2}. \label{eq:t-student}
\end{equation}
It is just a mere coincidence that the $\chi^2$ pdf which Beck chose to represent fluctuations
in the inverse temperature, (\ref{eq:beck}), has the same form as Maxwell's density for
the speed $\sqrt{\mbox{\small{$\frac{1}{n}$}}\sumn u_i^2}$. Quite surprisingly, the 
Student distribution (\ref{eq:t-student}) is independent of the variance, $1/\beta_0$.
Although this is important insofar as testing hypotheses of the mean of a population, because it
is independendent of the variance, is concerned, it nevertheless implies that information 
has been lost [\emph{cf\/}., (\ref{eq:in-beta-bis}) and 
(\ref{eq:F}) below]. \par 
The reason why Beck has $\beta_0$ as a parameter in his pdf (\ref{eq:Tsallis}) is that
he obtained 
\begin{eqnarray*}
\lefteqn{\frac{1}{\sqrt{2\pi}}\frac{1}{\Gamma(n/2)}
\left(\frac{n}{2\beta_0}\right)^{n/2}
\int_0^\infty\,\beta^{(n-1)/2}e^{-\half\beta(u^2+n/\beta_0)}\,d\beta}\\
& = & \frac{1}{B(\half,n/2)}\sqrt{\frac{\beta_0}{n}}\left(1+\frac{\beta_0u^2}{n}\right)^
{-(n+1)/2},\end{eqnarray*}
using (\ref{eq:beck}) instead of using (\ref{eq:chi}). Had he done so, he would have obtained
\[f_{u_0}(u)=\frac{1}{B(\half,n/2)}u_0^{-1}\left(1+\frac{u^2}{u_0^2}\right)^{-(n+1)/2}.\]
The two are the same if  equipartition  holds, 
\begin{equation}
u_0=\sqrt{\sumn u_i^2}=\sqrt{\frac{n}{\beta_0}}. \label{eq:equi-bis}
\end{equation}
\par
In general, the estimator of the variance, 
$\mbox{\small{$\frac{1}{n}$}}\sumn u_i^2$, will be different from the variance 
$\beta^{-1}_0$. The estimator is $\beta^{-1}$, and this temperature will usually be
different from the temperature of the reservoir, $\beta_0^{-1}$. In other words, the statistic for the Student
distribution, $t=u/\sqrt{\sumn u_i^2}$, is not distributed as a normal random variable,
as it would be had the statistic been given by  $t=u/\sqrt{n/\beta_0}$. The actual
variance can be completely unknown, and it suffices the sample variance alone in order
to make statistical predictions. Nevertheless, we expect, that as the number of degrees of freedom increases the
distribution of $t$ will be very similar to that of a standard normal variable 
(\emph{i.e.\/}, it is a consistent estimator). Accordingly,
we can identify
the $X_i^2$ in Beck's eqn (3), which are distributed as $\chi^2$, 
with the kinetic energies of the individual particles, and
the left-hand side should be $n/\beta$, and not $\beta$ itself.\par
The idea of \lq superstatistics\rq\ is that the exponential pdf
\cite{Beck03}
\begin{equation}
f_{\beta_0}(E)=\frac{e^{-\beta_0 E}\rho(E)}{Z(\beta_0)} \label{eq:exp}
\end{equation}
is actually a conditional pdf, and specifying a   \emph{normalized\/} pdf for what was
previously a mere parameter, say, by the $\chi^2$ pdf
\begin{equation}
f_{E_0}(\beta)=\frac{1}{\Gamma(n/2)}E_0^{n/2}\beta^{n/2-1}
e^{-\beta E_0}. \label{eq:beck-bis}
\end{equation}
leads to a new distribution of $E$ which is precisely the Tsallis pdf. However, 
(\ref{eq:beck-bis}) is no arbitrarily chosen pdf for the inverse temperature, but, rather
the  directing process which results when the parameter
of the original distribution is randomized. \par
Consider the usual case of a power law for the structure function \cite{Lavenda91},
\[
\rho(E)=\frac{E^{m-1}}{\Gamma(m)}, \label{eq:rho}
\]
which leads to equipartition  because the partition function is
$Z(\beta)=\beta^{-m}$. Then the process which is subordinated to the exponential pdf
(\ref{eq:exp}) has a pdf
\begin{equation}
f_{E_0}(E)=\int_0^\infty\,f_{\beta}(E)f_{E_0}(\beta)\,d\beta=
\frac{1}{B(m,n/2)}\frac{(E/E_0)^{m-1}}{(1+E/E_0)^{m+n/2}}\frac{1}{E_0}
\label{eq:in-beta-bis}
\end{equation}
which is still the inverted beta pdf [\emph{cf\/}., eqn (\ref{eq:in-beta})]. This is because the
L\'evy transform, 
\begin{equation}
\beta_0E=\beta E_0, \label{eq:levy-bis}
\end{equation} in (\ref{eq:exp}) produces a randomized inverse
temperature which is distributed according to (\ref{eq:beck-bis}) with $n=2m$ 
degrees of freedom. Consequently, the inverse beta pdf (\ref{eq:in-beta-bis}) is the 
Fisher-Snedecor pdf with equal degrees of freedom, $m=n/2$, or what is commonly known as
the Pareto distribution. The transformation to a
new process by randomizing the operational temperature cannot change the number of
degrees of freedom of the system. \par
\lq F-superstatistics\rq \cite{BC,Beck03} is not a directing process, but, rather, a 
subordinated process that is obtained from (\ref{eq:chi}) and (\ref{eq:n-maxwell}) by
averaging the product of these densities over a common value of the kinetic energy,
\emph{viz\/}. \cite{Lavenda95}
\begin{equation}
f_{\beta_0}(\beta)=\int_0^{\infty}\,f_{\beta_0}(u)f_{u}(\beta)\,du
=\frac{1}{B(n/2,n/2)}\frac{(\beta/\beta_0)^{n/2-1}}{(1+\beta/\beta_0)^n}
\frac{1}{\beta_0}.\label{eq:F}
\end{equation}
On account of the necessity that both the original and directing processes have the same
number of degrees of freedom, (\ref{eq:F}) is a symmetrical inverse beta distribution, 
which necessarily has \emph{the same
number of degrees of freedom\/}. Comparing (\ref{eq:F}) with (\ref{eq:in-beta-bis}),
it is easily seen that they are interchangeable on the strength of the L\'evy 
transformation (\ref{eq:levy-bis}). Reference to a Tsallis distribution in $\beta$ space
is completely inappropriate.\par
The symmetry of the inverse beta pdf (\ref{eq:F}) means that we are dealing with a 
composite system comprised of two subsystems with the same number of
degrees of freedom. The total energy $E_t=E_0+E$, formed from two subsystems with energies
$E$ and $E_0$, is fixed. The inverse beta pdf (\ref{eq:in-beta-bis}) is thus
 transformed into the beta pdf, which can be written as the composition law \cite[p. 80]
{Lavenda95}
\[\rho(E_t)=\int_0^{E_t}\,\rho(E_t-E)\rho(E)\,dE=\frac{E_t^{2m-1}}{\Gamma(2m)}\]
for the structure function.\par
Thus, we conclude that the statement 
\lq\lq Of course, other distribution functions
$f(\beta)$ can also be considered which may lead to other generalized statistics\rq\rq\
\cite{Beck02} is devoid of meaning when the original process is distributed according
to (\ref{eq:exp}). More importantly, as we have concluded elsewhere, \lq\lq
subordination can be considered as the probabilistic origin of power laws in physics
\rq\rq \cite[p. 58]{Lavenda95}.

\end{document}